# Enhanced Security of Public Key Encryption with Certified Deletion

Xiaogang CHENG[1], Ren GUO[2]

1. College of Computer Science and Technology, Huaqiao University, Xiamen 361021, China

2. College of Business Administration, Huaqiao University, Quanzhou 362021, China

**Abstract:** In classical cryptography, certified deletion is simply impossible. Since classical information can be copied any number of times easily. In quantum cryptography, certified deletion is possible because of theorems of quantum mechanics such as the quantum no-clone theorem, quantum superposition etc. In this paper, we show the PKE-CD (Public Key Encryption with Certified Deletion) scheme constructed in by Bartusek and Khurana in CRYPTO 2023 lack an important security property, which is important in practical applications. Then we show how to enhance this property, and construct a concrete scheme with this property. And we also discuss the relations between PKE-CD and other quantum cryptographic schemes such as quantum seal, quantum bit commitment etc.

## 1. Introduction

In classical cryptography, certified deletion [1] is simply impossible. Namely if Alice sends some classical information (whether encrypted or non-encrypted) to Bob, there is no way that Alice can be made sure that Bob have deleted the information. Since Bob can easily copy the classical information any number of times.

In quantum mechanics, there are some properties which are not possible in classical physics setting such as the no-cloning theorem, quantum superposition etc. So, in quantum cryptography these properties are exploited to achieve crypto system which are not possible in classical setting such as the unconditionally secure QKD (Quantum Key Distribution) [2,3], which are impossible in classical cryptography. Or to attack classical cryptographic schemes such as RSA, discrete logarithm via the famous Shor algorithm [4]. Other examples are quantum PSI (Private Set Intersection) [5,6], QSDC (Quantum Secret Direct Communication) [7,8], QD (Quantum Dialogue) [9], QSS (Quantum Secret Sharing) [10] and many others. QS (Quantum Seal) [11,12,13,14,15,16,17,18,19,20] is also a kind of cryptographic system which is simply impossible in classical setting. In QS, Alice can send a sealed information to Bob who are not supposed to open the seal until some predefined date or event. To check Bob's honesty, Alice can ask Bob to return the sealed information (which should be intact if Bob is honest) before the predefined date of event. Though it is possible to construct quantum seal scheme, it is shown that perfect QS scheme is impossible [20]. And there are fundamental constraints on the Bob's successful reading probability and the probability that Alice can find the cheating behavior. I.e., these two probabilities cannot be high at the same time. Later we show that there are closed relations between QS and certified deletion. Although quantum cryptography can achieve some crypto system which are not possible in classical setting, but there are also some fundamental no-go results for quantum seal [20], quantum bit commitment [21,22], quantum OTP (One-Time Programs) [23,24,25].

In this paper, we discuss how to use quantum cryptography to implement certified deletion, which is simply impossible in classical setting as mentioned above. In [], a concrete PKE-CD scheme is presented. While in this paper, we show that there is a shortage in the security property of the PKE-CD scheme in [1]. Namely, if the receiver of the PKE-CD is dishonest, he can actually steal the secret in the PKE-CD encryption and he can still generate legitimate certificate of deletion after this steal. Then we show how to enhance this security property with a concrete construction based on error correcting code and random quantum encoding. Though our construction is not perfect, but probabilistic, which means that Bob the receiver can only get the plain text information with 50% probability. We then show that probabilistic construction is actually inevitable, since perfect scheme is impossible. This is done by reducing the enhanced PKE-CD scheme to another well-known quantum cryptographic system quantum seal. We show that quantum seal and our enhanced PKE-CD schemes are reducible to each other. Since perfect QS is impossible, hence perfect enhanced PKE-CD is also impossible. Then we also discuss the relations between PKE-CD and other quantum cryptographic schemes such as quantum commitment and quantum OTP (One-Time Programs).

The paper is organized as following. In section 2, we review the original PKE-CD definition and construction of [1] and show the security shortage thereof. Then a concrete PKE-CD construction with enhanced security property is presented in section 3. In section 4, we analyze the security and relate PKE-CD with other well-known quantum schemes such as quantum seal, quantum bit commitment etc. And conclude in section 5.

**2. Review and security enhancement of the original PKE-CD scheme**

In this section, we briefly review the original PKE-CD definition, construction and security property in [1]. We show that the original security is not convenient in some practical applications, since Bob the receiver has to ask Alice the sender for secret keys for decryption. In our enhanced security definition, we remove this constraint, hence Bob can decrypt the ciphertext without the aid of Alice. And we also show that original PKE-CD construction in [1] fail to satisfy our enhanced security property.

**Definition 1. PKE-CD** is composed of the following algorithms:

1) GEN; 2) ENC; 3) DEC; 4) DEL; 5) VER

The first 3 algorithms are just normal post-quantum secure PKE schemes for generating public/secret keys, encryption and decryption. DEL is for generating certificate of deletion. And VER is for verifying the deletion certificate.

**Definition 2. Certified deletion security:** When input with the ciphertext, if the adversary (without the corresponding secret key) generates a legitimate deletion certificate. Then bit hidden in the PKE-CD encryption is information theoretically independent of the adversary's view.

In the security definition of PKE-CD in [1], the secret key is not in the possession of the receiver of the PKE encryption. I.e., if Bob the receiver wants to decrypt the encryption, he has to ask Alice the sender for the secret key. Otherwise, he cannot know the plaintext. In this paper, we want to relax this condition. In our scheme, Bob holds the corresponding secret key of PKE-CD, thus he can decrypt the PKE encryption to get the hidden information. But after this decryption, he will not be able to generate legitimate certificate of deletion. Hence Bob can directly decrypt the encryption with no more interaction with Alice the sender

**Definition 3. Refined certified deletion security:** With input of both the secret key and ciphertext, if the adversary generates and send back a legitimate deletion certificate which is guaranteed to pass Alice's verification, then bit hidden in the PKE-CD encryption is information theoretically independent of the adversary's view. And if the adversary tries to decrypt the ciphertext, with non-negligible probability he will get the hidden bit in the encryption. But then the adversary will not be able to generate a legitimate deletion certificate with non-negligible probability.

Next, we briefly review the original PKE-CD construction in [1] and show that the construction fails to satisfy the new security definition:

1. To encrypt a secret bit $b$, generate two random bit string with equal length $x$ and $\theta$.

2. Generate a quantum state $|x\rangle_\theta$, which means encode the bit $x_i$ with computational basis $\{|0\rangle, |1\rangle\}$ if the corresponding bit $\theta_i = 0$, with Hadamard basis $\{|+\rangle, |-\rangle\}$ if $\theta_i = 1$.

3. Use the post-quantum secure classical PKE scheme to encrypt the classical

information $\theta$ and $b \oplus \bigoplus_{x_i:\theta_i=0} x_i$, i.e.:

$$PKE(\theta, b \oplus \bigoplus_{x_i:\theta_i=0} x_i)$$

4. The final ciphertext are the PKE encryption and the quantum state $|x\rangle_\theta$ :

$$PKE\left(\theta, b \oplus \bigoplus_{x_i:\theta_i=0} x_i\right), |x\rangle_\theta$$

To attack this PKE-CD scheme, Bob can decrypt the encryption of $\theta$, and measure those bits $x_i$ of $x$, for which the corresponding $\theta_i$ is zero, by the computational basis. With these bits, the receiver Bob can easily decrypt and get the secret bit $b$.

Note that these measurements will not change the quantum state $|x\rangle$, since these qubits are encoded exactly by the computational basis. So, even after the decryption, the receiver can still provide a legitimate certificate of deletion, although he has already learned the secret, i.e. $b$.

We can compare this with the scheme of SS-CD (Secret Sharing with Certified Deletion). In $(2, 2)$ SS-CD, the two parties hold the following information respectively:

$$|x\rangle_\theta$$

$$ENC_{PK}(\theta, b \oplus \bigoplus_{x_i:\theta_i=0} x_i)$$

Again, if the two parties can collude then the certified deletion property is void. Since the party that hold $ENC_{PK}(\theta, b \oplus \bigoplus_{x_i:\theta_i=0} x_i)$ can decrypt to get $\theta$ and sent this information to the other party. Then as described above, with the information of $\theta$, the party that hold quantum state $|x\rangle_\theta$ can measure the corresponding qubits and get information about $\{x_i: \theta_i = 0\}$ without alerting the quantum state. Hence, he can still generate legitimate certificate of deletion. Even though the two parties have already gotten the secret. So, for this SS-CD scheme, it is essential that we should assume that the two parties should not collude to ensure the property of certified deletion. Without this assumption, again the certified deletion property is void. Note that this assumption makes no sense in the PKE-CD scheme, since in this scenario, both information (i.e. the quantum state and the basis used to generate it) are held by the same party, collusion is inevitable.

Let's just give a simple example. Let:

$$x = 1011\ 0110$$

$$\theta = 0011\ 1001$$

Then:

$$|x\rangle_\theta = |10++-11-\rangle$$

With the information of $\theta$, the ADV can measure the first, second, 6th and 7th qubit of $|x\rangle$ with the computational basis, and get 1011. Hence:

$$\bigoplus_{x_i:\theta_i=0} x_i = 1 \oplus 0 \oplus 1 \oplus 1 = 1$$

With this information, the secret bit b can be easily recovered. Note also that in this measurement, the quantum state $|x\rangle$ has not been changed. So later the ADV can still easily generate legitimate certificate of deletion even after this decryption. And the ADV can even throw away the quantum state $|x\rangle$, just keep classical info to generate legitimate certificate of deletion. The ADV can measure those qubits of |x> for which the corresponding bits of \theta are 1. In this case, 3rd, 4th, 5th and 8th qubits with Hadamard basis, ++-- will be gotten, namely 1100. Later he can use the following classical bit string as certificate of deletion:

$$**\ 11\ 0\ **\ 0$$

In which * just means random 0/1.

The problem above is that there is a way to measure the quantum state without changing it. So, to overcome this loophole, we must find a way to guarantee that the quantum state is measured and destroyed before the secret can be found. Hence if the secret is extracted, then there is no way to produce a legitimate certificate of deletion.

**3. Our PKE-CD construction with enhanced security**

Our PKE-CD construction with enhance security works as following:

1. By using Bob's public key, Alice encrypts a m-bit long string $B = b_1 b_2 \ldots b_m$ to ciphertext $C = c_1 c_2 \ldots c_k$. The PKE used are supposed to be post-quantum secure.

2. Choose a ECC (Error Correcting Code) with parameter $(n, k, d)$, which can encode a k-bit long message to a n-bit long code, and the minimum hamming distance is $d$. Hence up to $e = \lfloor \frac{d-1}{2} \rfloor$ bit errors can be corrected.

3. Use the ECC to encode the ciphertext $C = c_1 c_2 \ldots c_k$ to $D = d_1 d_2 \ldots d_n$.

4. Randomly choose the computational basis or Hadamard basis to turn $D = d_1 d_2 \ldots d_n$ into a quantum state $|D\rangle$ bit by bit.

5. Alice randomly choose $e$ places in the n-qubit long quantum state $|D\rangle$. And set each of these qubits to random value 0 or 1, by using computational or Hadamard basis randomly. Then send this quantum state with errors $|D'\rangle$ to Bob.

6. After receive the quantum state $|D'\rangle$, Bob can:

- Try to decrypt and get the message $b_1 b_2 \dots b_m$. He can randomly choose the computational basis or Hadamard basis, and measure each bit of $|D'\rangle$ with the chosen basis. If his choice is right, then he will get the classical message $D'$. Note there are might be $e$ errors, since the e places are random values as mentioned above. But these e errors can be corrected, since we use $(n, k, d)$ ECC. I.e., Bob can correct $D'$ to get $D$. Then decode $D$ to get the ciphertext $C$, and decrypt $C$ to get the plaintext message $B$. Note that the successful decryption probability is one half. With one half probability, Bob will fail to get the message if he chose the wrong basis to measure.
- Generate certificate of deletion. Bob can simply return the quantum state $|D'\rangle$ to Alice as deletion certificate. Of course, Alice can easily check if $|D'\rangle$ has been measured and destroyed, since she knows the basis used to encode each qubit of $|D'\rangle$.

Note that Bob can only choose one operation to perform. Namely if Bob chose to decrypt the encryption, then he will not be able to generate legitimate deletion certificate. Since he has to measure and destroy the quantum state. After which, he cannot reconstruct the quantum sate as certificate of deletion. If he chose to return the quantum state as deletion certificate to Alice, then he lost all the information about the plaintext.

**A toy example:**

Suppose the ciphertext length of the post-quantum PKE is 16 bits long. And we use the BCH code with parameters $(31,16,7)$, i.e. the length of the code is 31 bits long and the minimum Hamming distance is 7. Hence errors of up to $3 = \lfloor (7-1)/2 \rfloor$ bits are correctable.

1. Suppose the ciphertext of PKE is $C = 1000\ 1001\ 1010\ 1011$. Encode it with BCH (31,16,7):

$$D = 1000\ 1001\ 1010\ 1011\ \ 0010\ 0100\ 0101\ 000$$

2. Encoding the code above by the computational basis or Hadamard basis with equal probability. I.e., with 50% probability we use the $\{|0\rangle, |1\rangle\}$ basis:

$$|D\rangle = |1000\ 1001\ 1010\ 1011\ \ 0010\ 0100\ 0101\ 000\rangle$$

and 50% probability the Hadamard basis $\{|+\rangle, |-\rangle\}$:

$$|D\rangle = |+---+--++-+-+-++--+--+---+-+---\rangle$$

3. Randomly select 3 bits of the code, and send these 3 bits to random values 0 or 1, and encoding them also with random basis:

$$|\hat{0}\rangle \coloneqq \cos\left(\frac{\theta}{2}\right)|0\rangle + e^{i\psi}\sin\left(\frac{\theta}{2}\right)|1\rangle$$

$$|\hat{1}\rangle \coloneqq \sin\left(\frac{\theta}{2}\right)|0\rangle - e^{i\psi}\cos\left(\frac{\theta}{2}\right)|1\rangle$$

Namely, the azimuthal and polar angles are random in the range $(0,2\pi)$ and $(0,\pi)$ on the Bloch sphere respectively:

$\psi \in (0,2\pi), \ \theta \in (0,\pi)$

Suppose the three random places are 1, 3, 7, and the random values are 1,1,0, then the following quantum bits will be gotten:

$$|D'\rangle = |\hat{1}0\hat{1}0\ 10\hat{0}1\ 1010\ 1011\ \ 0010\ 0100\ 0101\ 000\rangle$$

$$|D'\rangle = |\hat{1}-\hat{1}-+-\hat{0}++-+-+-++--+--+---+-+---\rangle$$

4. Alice send these 31 qubits $|D'\rangle$ to Bob.

5. If Bob wants to decode and decrypt the encryption. He can measure the qubits by the computational or Hadamard basis. With 50% probability, he made the right choice which match the encoding basis that Alice used. Note that if Bob chooses the right basis, still he might get 3 wrong bits, since these 3 bits are deliberately set to random values by Alice in step 3. But that is not a problem, since ECC code is used, these 3 bits can be corrected by the decoding algorithm of the BCH $(31,16,7)$ code. Then Bob get the classical PKE ciphertext $C = 1000\ 1001\ 1010\ 1011$. With his own private key, he will be able to decrypt it to get the message $B$.

6. Instead of decoding and decrypting, Bob can return the unchanged quantum state $|D'\rangle$ to Alice as deletion certificate. Since Alice knows all the basis for encoding each qubit, she can easily check the validity of the certificate. The certificate can also be made classical. Alice can tell Bob what basis to measure for each qubit, all bases are random (i.e. random $\theta$ and $\psi$), except the bases for the random bits are the same with those encoding bases. Then Bob can measure $|D'\rangle$ with these bases, and return the result classical bit string to Alice. And Alice can check that all the bits are random, except that the error bits are the same as before.

**4. Security Analysis and relations with other cryptosystems**

In this section, we first discuss the security of our PKE-CD scheme, then we discuss the relationship between PKE-CD and other quantum cryptosystems such as quantum seal etc.

**Theorem 1.** Bob's successful decoding and decrypting probability is 50%. For malicious Bob, with 50% probability his cheating behavior will be detected by Alice.

**Proof**: For an honest receiver Bob, he simply randomly chooses a basis from computation or Hadamard basis to measure the quantum state $|D'\rangle$. Then clearly he will be able to decode and decrypt if his choice is the same with Alice, which happens with 1/2 probability. Whatever Bob's basis choice, after the measurement, he will lose the ability to generate a legitimate deletion certificate since the error qubits are destroyed. More precisely, if there are $e$ error bits, after measurement his probability of generating legitimate deletion certificate is less than:

$$\left(\frac{1}{2}\right)^e$$

It is negligible when $e$ is large.

Suppose a malicious Bob the receiver tries to steal the message without altering the quantum state $|D'\rangle$. His best strategy is to guess the basis used by Alice first, then try to measure the quantum state accordingly. Because as we know the computational basis and Hadamard basis are mutual unbiased. So, if Alice chooses computational basis to encode the message, most of the qubits would be $|0\rangle$ or $|1\rangle$, they will be basically indistinguishable (as in the QKD case) with those qubits if Alice choose the Hadamard basis. Hence Bob should first choose one from the computational and Hadamard basis. If Bob is honest, he will measure each qubit with the chosen basis. Then he will be able to decode and decrypt to get the plain text if his choice is right. Otherwise, the information encoded will be lost permanently. This bit-by-bit measurement operation of course will destroy the error bits which are encoded with random basis. Then later Bob will not be able to generate legitimate deletion certificate.

However, for malicious Bob, he may try other strategy as those in [20]. Instead of measuring each qubit one by one. He may choose to measure all the $n$ qubits of $|D'\rangle$ at the same time. And choose a POVM with $2^k$ projectors, each projector corresponds to a legitimate code with error bits less than or equal to $e$. Based on the outcome of this measurement, Bob can get the code, then decode and decrypt provided that his basis choice is right. Otherwise, he will also destroy the quantum state $|D'\rangle$ and lost the capability to generate legitimate deletion certificate.

**Relations with quantum seal and other quantum cryptosystems**

The PKE-CD scheme we constructed above is probabilistic, i.e. the probability of Bob's successful decryption is not 100% but 50%. Also, if Bob cheated, the probability that Alice will find this fact is not 100% percent. In fact, in this section we show that this is inevitable by reducing PKE-CD to another well-known quantum crypto scheme "Quantum Seal."

If we have a PKE-CD scheme with the new property we defined above, we can easily use it as a quantum seal scheme. Just send the cipher text of the PKE-CD scheme, then Alice can ask Bob to send back the certificate of deletion as a proof that Bob has not decrypt the cipher text.

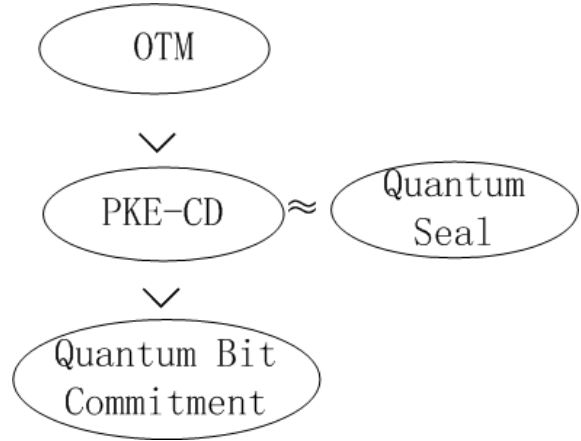

Fig 1. Known relations among some quantum cryptographic systems

Similarly, if we have a perfect seal scheme, then it can be used to construct PKE-CD as following. Just seal the ciphertext of a PKE encryption. The receiver Bob then can open the seal to get the PKE encryption, and decrypt with his own private key. Hence the difficulty of constructing PKE-CD and quantum seal scheme is almost the same. So, we have the scenario in Fig. 1. Quantum bit commitment is weaker than quantum seal and PKE-CD, while OTM is stronger [20].

Table 1. Comparison between our scheme with theoretical bounds

|  | $P_{reading}$ | $P_{dist}$ | $P_{NFP}$ |
|---|---|---|---|
| Upper bound[1] | $p$ | $1/2 + 1/4(2\sqrt{1-p} + 1 - p)$ | $1 - p^2 - \dfrac{(1-p)^2}{M-1}$ |
|  | 0.5 | 0.978553 | $\approx 0.75$ |
| Ours | 0.5 | 0.5 | 0.5 |

$P_{reading}$: the probability that Bob can decode and get the message.
$P_{dist}$: Alice's probability of determining whether Bob has cheated.
$P_{NFP}$: Alice's probability of Not-False Positive.

$M$: is number of message bits. In our case, we assume $M$ is large.

In Table 1, we compare our scheme with the theoretical upper bounds in [20]. We can see that our construction is well below the theoretical bounds in [20] in terms of Alice's distinguishing probability and probability of NFP. The upper bounds are 0.978533 and 0.75 respectively, while our scheme can only achieve 0.5 for both. Hence there are much rooms for improvement.

## 5. Conclusions

In this paper, we discuss the security of PKE-CD. We found that the original security definition of PKE-CD scheme is not good enough for some practical applications. Then we refine the security property of PKE-CD scheme.

Namely, in the original security definition of PKE-CD, the adversary does not know the secret key of the PKE in the PKE-CD scheme. This means that if the ciphertext receiver of PKE-CD want to decrypt the encryption, he has to ask the sender for the corresponding secret key. Otherwise, he cannot decrypt the encryption. Obviously, this is not very convenient in many practical scenarios. In our refined security definition, the adversary can hold the corresponding secret key. But under this new security definition, the original PKE-CD construction is completely insecure.

Then we also present a concrete construction which is secure under our new security definition. Though our construction is probabilistic, which means that the successful decryption probability is not one hundred percent, there are half chances that the decryption will fail. But then we show that this kind of probabilistic construction is inevitable since perfect PKE-CD scheme under our new security definition is impossible. This is done by reducing PKE-CD to other well-known quantum crypto schemes such as quantum seal, quantum bit commitment, quantum OTP etc.

In the future, we plan to improve our PKE-CD construction in terms of efficiency, security and successful probability etc. Theoretical bounds of PKE-CD and detailed relations with other quantum cryptographic schemes are also interesting research directions.